\documentclass[a4paper]{amsart}
  
\title[The Many Worlds of Modal $\lambda$-calculi: Part I]
{The Many Worlds of Modal $\lambda$-calculi: I. Curry-Howard for
Necessity, Possibility and Time}

\author{G. A. Kavvos}
\address{Department of Computer Science, University of Oxford.
Wolfson Building, Parks Road, Oxford OX1 3QD, United Kingdom}
\email{alex.kavvos@cs.ox.ac.uk}

\usepackage{prooftree}
\usepackage{epigraph}
\usepackage{natbib}
\usepackage{fancybox}
\usepackage{amssymb}
\usepackage{setspace}
\usepackage{hyperref}
\usepackage{mathtools}

\onehalfspace

\newcommand{\ctxt}[2]{#1\mathrel{;}#2}
\newcommand{\ibox}[1]{\mathsf{box\;}#1}
\newcommand{\letbox}[3]{\mathsf{let\;box\;} #1 \Leftarrow #2 \mathsf{\;in\;} #3}
\newcommand{\myeq}{\stackrel{\mathclap{\tiny\mbox{def}}}{=}}

\begin{document}

\maketitle

\tableofcontents

\epigraphwidth0.8\textwidth
\epigraph{Before embarking on details, here is one general piece
of advice. One often hears that modal (or some other) logic is
pointless because it can be translated into some simpler language
in a first-order way. Take no notice of such
arguments.}{Dana \cite{Scott1970}}

\pagebreak

\section*{Preface}

This is the first part of a threefold survey on the
subject of modal $\lambda$-calculi. Why threefold? We believe that
it is fair to say that work on constructive modal logic spans
three distinct yet intercommunicating streams:

\begin{description} \item[Curry-Howard] Work with a Curry-Howard
  concerns $\lambda$-calculi that arose as byproducts of the proof
  theory of modal logic, and their associated computational and
  categorical interpretations. This stream of work can be seen to
  date as far back as the 1950s, but witnessed rapid development
  during the 1990s, following the discovery of Girard's Linear
  Logic. Tracing the development of this material forms the
  subject of the first part of our survey.

  \item[Metaprogramming] It was sooner or later realized---as a
  result of the work on the Curry-Howard---that modal annotations
  can form the logical foundation that is much needed for all
  sorts of \emph{metaprogramming}, whether that comes in the form
  of binding-time analysis, staging of computations, or even
  partial evaluation. A flurry of developments on this stream
  happened in the early 2000s, and further work continues to
  appear in the 2010s.

  \item[Other applied calculi] A third stream of work, which
  constitutes a radical break from the proof-theoretic origins, is
  the use of modalities in all sorts of calculi that are
  fine-tuned to specific task. There are many possible such
  applications, ranging from security and depedency analysis to
  calculi for homomorphic encryption.
\end{description}

Whereas the first part of the survey is mainly the territory of
the present author, the task of surveying the metaprogramming
landscape---which should take up the second part of this
survey---is the domain of Mario Alvarez-Picallo
(\href{mailto:mario.alvarez-picallo@cs.ox.ac.uk}{mario.alvarez-picallo@cs.ox.ac.uk}).

\emph{We are actively on the lookout for a third co-author, in
order to cover the vast subject of applied calculi; please contact
us for more information.}

Finally, we also welcome any suggestions on comments on the
present material. The author can be reached by e-mail at
\href{mailto:alex.kavvos@cs.ox.ac.uk}{alex.kavvos@cs.ox.ac.uk}.

\pagebreak

\section{Introduction: Proof theory for modal logics}

\epigraphwidth0.8\textwidth
\epigraph{The situation has given rise to various suggestions. One
is that the Gentzen format, which works so well for
truth-functional operators, should not be expected to work for
intensional operators, which are far from truth-functions. (But
then Gentzen works well for intuitionistic logic which is not
truth-functional either.) Another suggestion is that the great
proliferation of modal logics is an epidemy from which modal logic
ought to be cured: Gentzen methods work for the important systems,
and the other should be abolished. `No wonder natural deduction
does not work for unnatural systems!'}{\cite{Bull1984}}

The history of modal proof theory is a long and tortuous one, and
not without good reason.  Attention shifted away from proof
systems for modal logic quite early in its development, for it was
not at all evident whether \emph{any} structural proof system was
appropriate. The blame was put on the `intensionality' of modal
logic, or the lack of appropriate symmetries. Indeed, these views
are explicitly stated by \cite{Bull1984} and echoed in the survey
of \cite{Negri2011}.  To make matters worse, attempting to combine
the Kripke semantics of intuitionistic logic with those of modal
logic leads to a bewildering variety of possibilities. 

We shall survey a stream of work on natural deduction systems for
a very specific class of normal modal logics, which has been
gradually handpicked over the years---the \emph{constructive modal
logics}. These share several characteristics: \begin{itemize}
  \item The modalities $\Box$ and $\lozenge$ are introduced
  independently, and not by the classic law $\Box A \equiv
  \lnot\lozenge\lnot A$ or its dual, as is common.  This is a
  welcome feature of intuitionistic analysis of logical
  connectives.

  \item They possess a variety of reasonably well-behaved proof
  systems. In fact, these logics have been selected over the years
  to admit analyses \emph{as close as possible} to the popular
  menu of proof-theoretic results: the cut-elimination theorem of
  \cite{Gentzen1935a, Gentzen1935b} for sequent calculi, and the
  normalisation theorem and subformula property of
  \cite{Prawitz1965} for natural deduction. Whether this goal has
  been achieved to a satisfying extent---or whether this is
  possible at all---has been the subject of debate for a long
  time.

  \item As a consequence of the striking similarity of the
  proof-theoretic analysis of these modal logic logics with that of
  intuitionistic propositional logic (\textsf{IPL}), they
  admit a \emph{Curry-Howard isomorphism} \citep{Curry1958,
  Howard1980}: each proof may be seen as a \emph{construction} for
  a \emph{type}, which is its formula. Consequently, we may read
  formulas as types, and proofs as programs. See, for example, the
  classic text of \cite{Girard1989}, or the modern treatment by
  \cite{Sorensen2006}. This leads to \emph{computational
  interpretations} of modalities as type formers, as well as
  associated \emph{modal $\lambda$-calculi}.

  \item Following the path, these logics all admit some sort of
  \emph{categorical semantics}, as pioneered for the typed
  $\lambda$-calculus and cartesian closed categories by
  \cite{Lambek1980}. These are invariably some kind of
  cartesian-closed category, coupled with a \emph{monoidal
  endofunctor} perhaps coupled with some additional categorical
  gadgets.

  \item In the same way that obtaining a constructive logic
  required eschewing the Law of the Excluded Middle
  (\textsf{LEM}), there are some modal principles that we have to
  eschew in order to obtain constructive modal
  logic. As in the case of \textsf{LEM}, the principles we have to
  reject are regarded as self-evident in traditional Kripkean
  analysis. These are usually \[
    \lozenge (A \vee B) \rightarrow \lozenge A \vee \lozenge B
      \quad\text{and}\quad
    \lnot\lozenge\bot
  \] or of a very similar form.
\end{itemize} It is fair to say that, even though the first
axiomatization of this kind is due to \cite{Wijesekera1990}, the
pioneers in adapting the Curry-Howard-Lambek isomorphism to modal
logic were \cite{Bierman1992a, Bierman1996a,
Bierman2000a}, and \cite{Moggi1989, Moggi1991}.

There are also other multiple other directions situated within the
intersection of intuitionism and modal logics.  One of the most
popular alternatives, mostly referred to simply as
\emph{intuitionistic modal logic}, was championed by Alex Simpson
in his thesis \citep{Simpson1994}. This approach is motivated by
Kripkean model theory, and indeed most of the effort is expended
in devising appropriate Kripke semantics for some modal logic
whose propositional component is intuitionistic. Invariably, the
proof systems considered by Simpson do not model \emph{validity},
but \emph{truth at a particular world}: judgments are of the form
$\Gamma \vdash A @ x$, which is intended to be read as ``under
assumptions $\Gamma$, A holds at world $x$.'' Propositional
assumptions aside, The context $\Gamma$ may also contain
assumptions of the form $x R y$, with the intended meaning that
``world $y$ is accessible from world $x$.'' \cite{Ranalter2010}
studies embeddings of the `constructive' systems into the
`intuitionistic' ones.

Even though there are very many sequent systems for intuitionistic
modal logic, we shall not survey sequent calculi here, despite
their close relationship to natural deduction systems. There is a
wide variety of those, ranging from austere Gentzen-style to
far more radical proposals. The former were pioneered by
\cite{Ohnisi1957, Ohnisi1959}, and are surveyed by \cite{Ono1998}.
\cite{Wansing2002} surveys both ordinary and `generalized' sequent
systems. \cite{Negri2011} adopts an even wider viewpoint, also
covering work on some `non-traditional' formalizations. A broader
survey of the intersection of intuitionism, and modal logic,
including applications to computer science, is the editorial
\citep{DePaiva2004}, which is followed up by \citep{Stewart2015}.

Returning to our present concerns, we may list the following
systems that fit our aforementioned criteria: \begin{itemize}
  \item a constructive version of the Lewis system $\mathsf{S4}$, which is
  known $\mathsf{CS4}$; 

  \item a constructive version of the smallest
  normal modal logic $\mathsf{K}$, which is known as $\mathsf{CK}$;

  \item constructive versions of the logic $\mathsf{K4}$ and the
  logic of provability $\mathsf{GL}$, both entirely due to
  \cite{Bellin1985}; and

  \item \emph{propositional lax logic} \textsf{PLL},
  also known as \emph{computational logic}, denoted \textsf{CL}.

  \item \emph{constructive linear temporal logic}, \textsf{CLTL}
\end{itemize} 

Of these, the logics \textsf{CS4} and \textsf{PLL}/\textsf{CL}
have been the subject of intense scrutiny. The proof theory of
\textsf{CS4} has been a popular subject, ever since the days of
\cite{Prawitz1965}. Its categorical interpretation reveals that
there is a latent \emph{universal property} underlying it, which
makes for rather well-behaved proof systems, with plenty of
harmony. On the other hand, \textsf{PLL}/\textsf{CL} is the logic
underlying Moggi's \emph{monadic metalanguage}, which has
enjoyed considerable attention as (a) a popular way to structure
denotational semantics, (b) a way to encapsulate impurities and
effects in functional languages, and (c) a pattern in which many
programs can be structured, in a modular way. As such, the
underlying logic has generated some interest, even if it was
originally discovered and discarded by \cite{Curry1952}, mainly
because of its unusual properties.

The logic \textsf{CK}, a minimal modal extension of the
intuitionistic propositional logic, has received very little
attention. It was devised as a fragment of proof systems for
\textsf{K4} and \textsf{GL} by \cite{Bellin1985}, whose other
proof systems have not been studied since their publication.

Finally, \textsf{CLTL} is a very simple constructive fragment of
classical \emph{Linear Temporal Logic}, involving only the `next'
operator. Notwithstanding its simplicity, it has numerous
applications in \emph{metaprogramming}.

\subsection{Preliminaries}

All but two of our modal logics $\mathcal{L}$ shall be sets of
formulae---the \emph{theorems} of the logic. These formulae are
generated by the following Backus-Naur form: \[
  A, B \quad ::=\quad P
    \;|\; \bot
    \;|\; A \land B
    \;|\; A \lor B
    \;|\; A \rightarrow B 
    \;|\; \Box A
    \;|\; \lozenge A
\] where $P$ is drawn from a countable set of propositions. The
two exceptions to this rule replace the cases of $\Box A$ and
$\lozenge A$ by a single modality, $\bigcirc A$.

The sets of theorems will be generated by \emph{axioms}, closed
under some \emph{inference rules}. The set of axioms will always
contain a complete axiomatization of intuitionistic propositional
logic (\textsf{IPL}). The set of inference rules will consist of
the two rules necessary to capture \textsf{IPL}, namely the
\emph{axiom rule}: \[ \begin{prooftree}
  A \text{ is an axiom }
    \justifies
  \vdash A
\end{prooftree} \] and the rule of \emph{modus ponens}: \[ \begin{prooftree}
  \vdash A \rightarrow B \quad \vdash A
    \justifies
  \vdash B
\end{prooftree} \] As for the modal part, in all but two cases, we
shall employ the necessitation rule, namely \[ \begin{prooftree}
  \vdash A
    \justifies
  \vdash \Box A
\end{prooftree} \] Finally, any logic $\mathcal{L}$ will be closed
under substitution of theorems for propositional constants.

\section{Constructive S4}
  \label{sec:cs4}

The founding stone of constructive modal logics has been the
constructive logic underlying the Lewis system $\mathsf{S4}$
\citep{Lewis1932}. A simple axiomatization of $\mathsf{CS4}$, due
to \cite{Alechina2001}, comprises the axioms of \textsf{IPL}, and
also \begin{align*}
  &(\mathsf{K)} & \Box (A \rightarrow B) \rightarrow (\Box A
  \rightarrow \Box B) \\
  &(\mathsf{4}) & \Box A \rightarrow \Box\Box A \\
  &(\mathsf{T}) & \Box A \rightarrow A
\end{align*} To obtain the diamond fragment, we add \begin{align*}
  &(\lozenge\mathsf{K}) & \Box(A \rightarrow B) \rightarrow
    (\lozenge A \rightarrow \lozenge B) \\
  &(\lozenge\mathsf{4}) & \lozenge\lozenge A \rightarrow \lozenge A \\
  &(\lozenge\mathsf{T}) & A \rightarrow \lozenge A
\end{align*}

Studies of the proof theory of $\mathsf{CS4}$ abound, making it
the most widely studied constructive modal logic. The fundamental
reason for this success can only be grasped once one interprets
its axioms categorically: take $\Box$ to be a \emph{endofunctor}
on a cartesian closed category. To interpret axiom \textsf{K}
then, we need a natural transformation \[
  m_{A, B} : \Box A \times \Box B \rightarrow \Box (A \times B)
\] which would make the functor \emph{monoidal}. Then, to
interpret axiom \textsf{T} we need \[
  \epsilon : \Box \Rightarrow (-)
\] and to interpret axiom \textsf{4} we need \[
  \delta : \Box \Rightarrow \Box^2
\] These contraptions look like they could take the place of
counit and comultiplication of a \emph{monoidal comonad}. As such,
they give rise to an \emph{adjunction}, either through the
Eilenberg-Moore or Kleisli constructions (see e.g.
\cite{Awodey2010}).  Thus, there is a latent \emph{universal
property}, which lends the proof systems of \textsf{CS4}
reasonable symmetry, from which one can mould introduction and
elimination rules.

The Kripke and algebraic semantics for \textsf{CS4} have been
thoroughly investigated by \cite{Alechina2001}.

\subsection{Early attempts}

The first attempt at modal natural deduction was the system for
the ($\Box$) fragment of \textsf{S4} and \textsf{S5} defined and
studied by \cite{Prawitz1965}. A rule for diamond is also given,
but was neither analyzed nor discussed. Prawitz considers
the obvious solution: if all assumptions are modalized, then we
are allowed to put a box in front of the conclusion. In
sequent-style notation, \[ \begin{prooftree}
  \Box\Gamma \vdash A
    \justifies
  \Box\Gamma \vdash \Box A
\end{prooftree} \] where \[
  \Box(A_1, \dots, A_n) := \Box A_1, \dots, \Box A_n
\] A system for \textsf{S5} is also obtained if assumptions are
either modalized, or the negation of a modalized formula.

Prawitz' systems are thoroughly criticized in \citep{Bierman2000a}.
Their main defect are that they are \emph{not closed under
substitution}: if we substitute any derivation of $\Box A_i$ which
has open non-modal assumptions at the top, then we do not
necessarily get a derivation whose final step is not valid: it
is no longer the case that all assumptions are modalized. 

Here is a classic example of this phenomenon, taken from \emph{op.
cit.} Suppose that we have a derivation of the form \[
  \begin{prooftree}
  \[ \[ \leadsto \]  \justifies \Box A \vdash B \]
    \justifies
  \Box A \vdash \Box B
\end{prooftree} \] Letting $\Gamma = C \rightarrow \Box A, A$, we
derive \[
\begin{prooftree}
  \Gamma \vdash C \rightarrow \Box A
    \quad
  \Gamma \vdash C
    \justifies
  \Gamma \vdash \Box A
\end{prooftree} \] Substituing, now, this derivation of $\Box$ for
the assumption in $\Box A \vdash \Box B$, we get that \[
\begin{prooftree}
  \[ \Gamma \vdash C \rightarrow \Box A
      \quad
    \Gamma \vdash C
      \justifies
    \Gamma \vdash \Box A \]
      \quad
  \[ \[ \[ \leadsto \]  \justifies \Box A \vdash B \]
    \justifies
  \Box A \vdash \Box B \]
    \justifies
  \Gamma \vdash \Box B
\end{prooftree} \] Eliminating the detour, we would have \[
\begin{prooftree}
  \[ \[ \leadsto \] \justifies \Gamma \vdash B \]
  \justifies
    \Gamma \vdash \Box B
\end{prooftree} \] which is not a valid application of the
introduction rule at all: $A$ bears no box in $\Gamma$.

The inherent difficulties of this system were suspected by Prawitz,
mainly because there are roundabout ways to derive formulas, i.e.
ineliminable detours. In \citep[\S VI.2]{Prawitz1965}, he
introduces a weaker restriction by the name of \emph{essentially modal
formulas}. However, it is noted in \citep[\S 8]{Bierman2000a} that
this system eliminates \emph{too many detours}: categorically
speaking, it forces $\Box A \cong \Box\Box A$.

Some more early work, none of which explicitly attributed to
Prawitz, is surveyed by \cite{Satre1972}. Prawitz' rule appears as
$\Box I_4$ and is attributed to \cite{Ohnisi1957, Ohnisi1959}, but
its proof-theoretic significance is not discussed.

\subsection{The Experience of Linear Logic}

Things took a radical turn with the discovery of Linear Logic by
\cite{Girard1987}. The explicit control of assumptions\footnote{It
is interesting to remind the reader that controlling weakening and
contraction of assumptions was an early and now-defunct goal of
modal logics, through the notion of \emph{strict implication}; see
\citep[\S II.2]{Girard1987}.} seemed to pave the way for a
multitude of applications in logic and computation.

One of the first goals of the community was to isolate an
natural deduction system for the smallest fragment of Linear
Logic, namely Intuitionistic Linear Logic (denoted
\textsf{ILL}).\footnote{Girard's presentation was in terms of an
expertly symmetric sequent calculus for Classical Linear Logic}
This would amount to a very basic \emph{linear
$\lambda$-calculus}, which would occupy central ground in numerous
applications. It so happened that the advancements made in this
field had a significant impact on the study of natural deduction
for modal logics, and so we rapidly survey them here.

An attempt to produce a system of categorical combinators by
\cite{Girard1986} and \cite{Lafont1988} led to the first
formulation of natural deduction for \textsf{ILL} by
\cite{Abramsky1990, Abramsky1993}. Even though it served
its original purpose, Abramsky's treatment of the ``of course''
($!$) modality was similar to Prawitz' for \textsf{S4}, and hence
\emph{not closed under substitution}. A counterexample was
discovered by \cite{Wadler1991, Wadler1992, Wadler1994}, who also
realized that whenever substitution accidentally leads to a valid
deduction, then soundness of the obvious interpretation in the
categorical semantics of \cite{Seely1989} necessitates that $!A
\cong\ !A\ \otimes\ !A$.

The challenge was immediately taken up by \cite{Bierman1992} and
\cite{Benton1993b}, whose treatment led to a rather complicated
system: as their formulation of natural deduction was based on
generalized rules in the style \cite{Schroeder-Heister1984}, the
$\lambda$-terms of their language contained \emph{explicit
substitutions} \citep{Abadi1991}. This system was very influential
in subsequent developments for modal logic.

Alternative syntax for \textsf{ILL} was also proposed by
\cite{Troelstra1995}, the system of whom is based on adapting
Prawitz' `essentially modal formulae' restriction from \textsf{S4} to
\textsf{ILL}; this solves the problem with substitution, but also forces
${!}A \cong {!}A \otimes {!}A$.

A further breakthrough occurred when the style of Girard's Logic
of Unity \citep{Girard1993} was distilled by \cite{Andreoli1992}
and \cite{Wadler1993, Wadler1994} into a new syntax involving two
contexts---one for `linear' and one for `intuitionistic'
(``banged'') variables. An even simpler system of the same kind
was rediscovered by \cite{Plotkin1993}, and extensively studied by
his student \cite{Barber1996, Barber1997}. This logic is known as
\textsf{DILL} (Dual Intuitionistic Linear Logic), and still
comprises one of the two state-of-the-art systems for
\textsf{ILL}.

Another kind of system was produced \cite{Benton1994}, which
involves two different types of sequent: intuitionistic and
linear. Benton's system offers very convenient view of the
semantics of \textsf{ILL}---for which see \cite{Mellies2009}---as a
\emph{linear-non-linear} adjunction. This is also another type of
system which we may view as the state of the art, and often the
initial point of further study. Even though this style of system
is found in more programming applications nowadays, it does not
appear to have exerted any influence at all on the study of
modalities.

\subsection{The Bierman \& de Paiva system}

Very soon after contributing to the \textsf{ILL} system in
\cite{Bierman1992} and \cite{Benton1993b}, \cite{Bierman1992a,
Bierman1996a, Bierman2000a} noticed that the rules for the ``of
course'' ($!$) modality of \textsf{ILL} are really the same as those for
\textsf{S4}. It was an easy step to remove the structural
restrictions of \textsf{ILL}, in order to uncover a natural deduction and
term assignment system for \textsf{CS4}.

The approach employed by \cite{Benton1993b} for \textsf{ILL}, and exploited
by \cite{Bierman1992a} for \textsf{CS4} is based on the same
underlying principle: \begin{center}
  \shadowbox{
    \begin{minipage}{0.7\textwidth}
    \begin{center}If the \emph{modal substitution (cut) rule} is
    not admissible, then ``bolt it'' onto the introduction
    rule. \end{center}
    \end{minipage}
  }
\end{center} That is: if the strategy of requiring that all
assumptions are modalized in order to infer a modal formula breaks
substitution, then simply include a side condition of the form
``\dots and, by the way, we can fill in some deduction for these
modalized formulae already.'' Hence their introduction rule: \[
\begin{prooftree}
  \Gamma \vdash M_1 : \Box A_1 
    \quad \dots \quad
  \Gamma \vdash M_n : \Box A_n
    \quad\quad
  x_1 : \Box A_1, \dots, x_n : \Box A_n \vdash N : B
    \justifies
  \Gamma \vdash \textsf{box } N \textsf{ with } M_1, \dots, M_n
  \textsf{ for } x_1, \dots x_n : \Box B
\end{prooftree} \] Except for the fact it works, there is almost
nothing good about this rule. It appears to be an introduction
rule, yet the connective it introduces also appears in the
conclusion of the (minor) premises, so it is also a cut rule! This
is coupled with the obvious way to eliminate boxes, namely \[
\begin{prooftree}
  \Gamma \vdash M : \Box A
    \justifies
  \Gamma \vdash \textsf{unbox } M : A
\end{prooftree} \] which corresponds to the axiom $\Box A \rightarrow
A$. The associated $\beta$-rule is \[
  \textsf{unbox }(\textsf{box } N \textsf{ with } M_1, \dots, M_n
    \textsf{ for } x_1, \dots x_n) \longrightarrow
    N[M_1/x_1, \dots, M_n/x_n]
\] Thus, to modalize a formula, (1) \emph{we need to be able to
prove it only on modal assumptions}, and (2) \emph{we must be able
to fullfil all of them}. Their proofs are then captured as part of
the proof term, which, in the spirit of \cite{Abadi1991},
constitute an \emph{explicit substitution}. The explicit
substutitions are then effected en masse when the proof term is
`unboxed' (cf. evaluated, interpreted).

Contrary to what is stated in \citep{Bierman2000a}, this
introduction rule for box does \emph{not} appear in
\citep{Satre1972}, and is their own invention. The rest of
\citep{Bierman2000a} is dedicated to the study of the resulting
natural deduction system and $\lambda$-calculus, as well as its
categorical semantics.

The original system of \cite{Bierman1992a, Bierman1996a} was
extended by \cite{Kobayashi1997} with a rule for diamond, which
was then incorporated into the journal version of their article
\citep{Bierman2000a}. We may introduce a diamond at any time: \[
\begin{prooftree}
  \Gamma \vdash M : A
    \justifies
  \Gamma \vdash \textsf{dia } M : \lozenge A
\end{prooftree} \] Eliminating a diamond can \emph{only} be done
in conjunction with a bunch of boxed assumptions. Namely, if we
use exactly some boxed variables $x_i : \Box A_i$ and a single
variable $y : B$ to produce a term of diamond type, then, if we
fulfill all the boxed assumption and provide something of diamond
type for the $y : B$ variable, we can indeed produce the term of
diamond type: \[
\begin{prooftree}
  \Gamma \vdash M_1 : \Box A_1 
    \quad \dots \quad
  \Gamma \vdash N : \lozenge B
    \quad\quad
  x_1 : \Box A_1, \dots, x_n : \Box A_n, y : B \vdash N : \lozenge C
    \justifies
  \Gamma \vdash
    \textsf{ let dia } y = N \textsf{ in } N \textsf{ with }
      M_1, \dots, M_n \textsf{ for } x_1, \dots, x_n : \lozenge C
\end{prooftree} \] with the reduction rule \begin{align*}
  &\textsf{ let dia } y = \textsf{dia } P \textsf{ in } N \textsf{ with }
    M_1, \dots, M_n \textsf{ for } x_1, \dots, x_n \\
  &\longrightarrow N[\vec{M_i}/\vec{x_i}, P/y]
\end{align*}

Putting the lack of `good symmetries' aside, there are deeper
flaws in the aforementioned rules.  Notably, a large number of
\emph{commutative conversions} are needed to expose `hidden'
redices, which spoil the so called \emph{subformula property}. As
an example of this phenomenon, quoted by \cite{Pfenning1995},
consider the term \[
  \textsf{box } (\textsf{unbox}\ f\ c) \textsf{
  with } (\textsf{box } (\lambda x : A. x)) \textsf{ for } f
\] and notice that it has `hidden' an obvious redex by `splitting
it' between the term and the explicit substitution. The subformula
property breaks, and this detour cannot be eliminated---not even
in conversion, let alone in reduction.

The issue of commutative conversions is known to arise from rules
for positive connectives in Natural Deduction, such as those for
disjunction ($\lor\mathcal{I}$) and existence ($\exists
\mathcal{I}$)---for a particularly perspicuous discussion, see
\cite[\S 10.1]{Girard1989}. These rules are very similar to rules
in the style of \cite{Schroeder-Heister1984}, which in turn are
very similar to those of Bierman and de Paiva's system.

In designing their system for \textsf{ILL}, \cite{Bierman1992} and
\cite{Benton1993b} included some of commutative conversions in
their definition of reduction, whilst only recovering the complete
set in the categorical interpretation. However, they do not
investigate the subformula propert at all. Similarly,
\cite{Bierman2000a} do present most commutative conversions as
conversions between natural deduction proofs, but do not push them
down to the term level. As such, they prove neither the subformula
property nor completeness of the categorical semantics.

These issues were resolved by \cite{Goubault-Larrecq1996}, who
discusses the present system and systematically remedies its
shortcomings. In the process, he provides full proofs of all
common logical properties, but without considering categorical
models. Goubault-Larrecq remarks that the introduction of explicit
substitutions necessitates the addition of further
\emph{structural rules} concerning them. For example, the term \[
  \ibox{\lambda u. uyz} \textsf{ with } x, x \textsf{ for } y, z
\] does not correspond to a cut-free proof, but the ability to
\emph{contract} the two $x$'s to \[
  \ibox{\lambda u. uyy} \textsf{ with } x \textsf{ for } y
\] does. Similarly, the non-occurence of variables that have been
explicitly substituted for, i.e. \[
  \ibox{\lambda x. x} \textsf{ with } M \textsf{ for } y
\] should really be \emph{weakened} to \[
  \ibox{\lambda x. x} \textsf{ with } \langle\rangle \textsf{ for
  } \langle\rangle
\] All is well with these rules, which explicitly mirror
\emph{contraction} and \emph{weakening} on the level of explicit
substitutions, and indeed they are mentioned by
\cite{Bierman2000a} as well. Surely, we also need \emph{exchange},
which would provide us with \[
  \ibox{M} \textsf{ with } P, Q \textsf{ for } x, y =_\beta
    \ibox{M} \textsf{ with } {Q, P} \textsf{ for } y, x
\] which he also includes. He then proceeds to argue that these
rules, whilst necessary, they jeopardize the computational meaning
of the system.  Indeed, these are rules traditionally found as
\emph{structural rules} in sequent calculi, but in natural
deduction they are usually admissible.

To highlight why the above issues constitute an appreciable
shortcoming, we need to recall that, in proof theory, it is
natural deduction proofs that are thought to comprise the ``real
proof objects.'' In contrast, sequent calculi allow for more
symmetry precisely so they can facilitate proof-theoretic
analysis. To quote \cite[\S 5.4]{Girard1989}: \begin{quote}
  ``The translation from sequent calculus into natural deduction is
  not 1-1: different proofs give the same deduction, [...]

  In some sense, we should think of the natural deduction rules as
  the true ``proof'' objects. The sequent calculus is only a
  system which enable [sic] us to work on these objects [...]

  In other words, the system of sequents is not primitive, and the
  rules of the calculus are in fact more or less complex
  combinations of rules of natural deduction [...]''
\end{quote} It follows, argues Goubault-Larrecq, that the Bierman
and de Paiva calculus does \emph{not} expose the computational
behaviour of S4 proofs: if it did, it would need no structural
rules at all.

\subsection{The Goubault-Larrecq system}

To mitigate the deficit introduced by the extra structural rules
in Bierman and de Paiva's system,
\cite{Goubault-Larrecq1996, Goubault-Larrecq1996a,
Goubault-Larrecq1996b, Goubault-Larrecq1997} proceeds to define
his own system, namely the $\lambda\textsf{ev}Q$ calculus.
The calculus itself is a very peculiar mixture of
\emph{categorical combinators} \citep{Curien1993} and
\emph{explicit substitutions} \citep{Abadi1991}. He then proceeds
to exhaustively study it, in four manuscripts of epic proportions.

One of Goubault-Larrecq's fundamental contributions is that, in
\cite[\S 2]{Goubault-Larrecq1996} and \cite[\S
2.3]{Goubault-Larrecq1996a}, he is the first to draw attention to
the similarity of the computational behaviour of \textsf{CS4} to
the \emph{quoting} features of LISP (see \cite{Bawden1999}) and
the \emph{reflective towers} of \cite{Smith1984} (see also the
discussion in \cite{Goubault-Larrecq1997a}. Even though the
connection with staged metaprogramming was thrust into the
limelight by \cite{Davies2001a}---see \S \ref{sec:dcs4}---this
particular line of thought has not been subjected to further
investigation, save from one comment in \emph{op. cit.}, who
remark that the behaviour of their modal language ``is closer to
the quotations of a ``semantically rationalized dialect'' of Lisp,
called 2-Lisp,'' referring to the intermediate, `non-reflective'
stepping stone for Smith's reflective towers \citep{Smith1984}.
However, it seems that the connection between constructive linear
temporal logic and metaprogramming in LISP is closer---see
\S \ref{sec:cltl}.

\subsection{The Martini \& Masini, Pfenning \& Wong systems}

Following previous work by \cite{Masini1992, Masini1993} on
\emph{2-sequent calculi} and an associated system of natural
deduction, \cite{Martini1995} present and study a system of
\emph{intuitionistic 2-sequents} and an associated \emph{levelled
$\lambda$-calculus} that, with minor structural variations, which
seemingly captures all three logics \textsf{CK}, \textsf{CK4}, and
\textsf{CS4} (this latter correspondence they do not prove).
In some ways, their system is elegant and crisp; but the
metatheory is difficult, and the levelling on the
$\lambda$-calculus is complicated. No treatment of the $\lozenge$
modality is provided.

Drawing impetus from the above presentation, \cite{Pfenning1995}
``flatten out'' its two-dimensional structure into a \emph{stack
of contexts}. This provides them with a more workable calculus
based on the same idea, and allows them to easily prove various
meta-theoretic properties.  For illustration, we use the improved
version found in \cite[\S 5]{Davies2001a}. A sequent has the form
\[
  \Psi ; \Gamma  \vdash M : A
\] where $\Psi$ is a list of contexts, the \emph{context stack}.
The idea is that, every $\Gamma_i$ in $\Psi = \Gamma_1, \dots,
\Gamma_n$ represents an `arbitrary world' reachable from
$\Gamma_{i-1}$.  The word `world' in this context is supposed to
invoke the look and feel of Kripke semantics \citep{Kripke1963},
but the idea is more subtle, and no actual accessibility relation
is involved. 

A context stack of the form \[
  \Psi ;\ \Gamma ;\ \cdot
\] (where $\cdot$ is the empty context) is meant to be thought of
as a `path' through $\Psi$, leading to the world $\Gamma$, and then
into some arbitrary world reachable from $\Gamma$.  So, if
something \emph{holds} at the end of this path, i.e. \[
  \Psi ; \Gamma ; \cdot \vdash M : A
\] then it follows that the same thing holds \emph{at every world
we can reach from $\Gamma$}. In the language of Kripke, $\Box A$
holds at $\Gamma$: \[
  \Psi ; \Gamma \vdash \textsf{box } M : \Box A
\] This is the introduction rule. Of course, we can only use
variables from the current world, i.e. the variable rule is \[
  \Psi ; \Gamma, x{:}A, \Gamma' \vdash x : A
\] The fundamental tenet is the following: \begin{center}
  \shadowbox{
    \begin{minipage}{0.8\textwidth}
    \begin{center}
      Force terms of type $\Box A$ to be \emph{closed} with
      respect to any world accessible from the current one, but
      allow them to use whatever is available in the `current
      world' ($\Gamma$), or a world visited in the path to the
      current one ($\Psi$).
    \end{center}
  \end{minipage}}
\end{center} The elimination rule, which allows us to `push' boxed
terms in some world forwards, in order to recover them in any
`accessible' world. For example, \[
\begin{prooftree}
  \cdot\ ; x : \Box A \vdash x : \Box A
    \justifies
  x : \Box A ; \Gamma \vdash \textsf{unbox}_1\ x : A
\end{prooftree} \] is a way to push the variable $x$ to any
arbitrary accessible world.  In full generality, if we have a
term $\Psi ; \Gamma \vdash M : \Box A$ then we can push
the it to whichever accessible world we desire: \[
  \Psi ; \Gamma ; \Gamma_1 ; \dots ; \Gamma_n \vdash
    \textsf{unbox}_n\ M : A
\] no matter how far in the `future' ($\Gamma_1, \dots, \Gamma_n$)
we go. The corresponding $\beta$-rule is \[
  \textsf{unbox}_n\ (\textsf{box } M) \longrightarrow
    \{n/1\}M
\] where $\{n/p\}$ is a relatively involved operation on syntax
which relabels occurrences of $\textsf{unbox}_n$ in a consistent
way.

The notion of `accessible world' is very flexible, in that restricting
it allows us to limit the system to the logics \textsf{CK} or
\textsf{CK4}--see \S \ref{sec:ck}, \S \ref{sec:ck4gl}. But this is
a system for \textsf{CS4}, and if---as in the Kripke semantics for
\textsf{S4}---our accessibility relation is \emph{reflexive}, then
\emph{the present is accessible}, and we may as well
`unbox'/evaluate anything we have: \[
  \begin{prooftree}
    \Psi; \Gamma \vdash M : \Box A
      \justifies
    \Psi; \Gamma \vdash \textsf{unbox}_0\ M : A
  \end{prooftree}
\] This corresponds to \emph{axiom T}: $\Box A \rightarrow A$. And
if our accessibility relation is \emph{transitive}, we allow all
constructs $\textsf{unbox}_n$, for $n = 1, 2, \dots$, so we may
push anything as far down an accessible chain of worlds, obtaining
a version of the logic \textsf{CK4} with \emph{axiom 4}. Even
further, if we limit ourselves to the case $n = 1$, we get simply
\textsf{CK}.  This is an observation that neither
\cite{Pfenning1995} nor \cite{Davies2001a} capitalized on even
though they were the first to make it.

\subsection{The Davies \& Pfenning systems}
\label{sec:dcs4}

\cite{Davies1996, Davies1999, Davies2001a} suggested that one can
intuitively read values of type $\Box A$ as ``code of type $A$,''
in an interpretation that the present author calls
\emph{modality-as-intension}. The fundamental idea was first aired
in \citep{Pfenning1995}, and is this: if we interpret box as an
\emph{intensional type}, then programming in that type theory
really corresponds to \emph{metaprogramming}: manipulating terms
at modal types exactly corresponds to manipulating code, in a
logical and type-safe way.

\cite{Davies2001a} present two systems for \textsf{CS4}: the
leaner version of the system of \cite{Pfenning1995} that we have
just described (therein called the `implicit' formulation), and a
two-context calculus (the `explicit' system), which is eerily
similar to what one were to obtain obtain if they were to remove
linear restrictions from the system \textsf{DILL} of
\cite{Barber1996} and \cite{Plotkin1993}. This is, however,
independent of the work of Barber and Plotkin, but shares common
ancestry with it in the calculi of \cite{Girard1993},
\cite{Andreoli1992}, and \cite{Wadler1993, Wadler1994}. We call
the dual-context calculus of Davies and Pfenning by the name
\textsf{DCS4} (`dual constructive \textsf{S4}').

Translations are provided between the `implicit' and `explicit'
version, along with proofs of correctness. The connection with
metaprogramming comes from an embedding of \emph{two-level
$\lambda$-calculus} of \cite{Nielson1992} in the `implicit
system.' The two-level $\lambda$-calculus, descended from an
untyped predecessor devised by \cite{Gomard1991}, is a language
where expressions are annotated (and type-checked) as being
available in two stages: either at \emph{compile-time}, or
\emph{run-time}. This is done in such a way that that compile-time
expressions never depend on run-time expressions and can be
evaluated in advance. The contribution of \cite{Davies1996,
Davies1999, Davies2001a} is the demonstration that the two-level
calculus can be embedded in a fragment of their
`implicit'/context-stack \textsf{CS4} system, which corresponds to
the logic \textsf{CK}. This led to a flurry of developments in
calculi for metaprogramming, and is the beginning point of the
second part of this survey; see also \S \ref{sec:cltl}.

Sequents of \textsf{DCS4} have the form \[ 
  \ctxt{\Delta}{\Gamma} \vdash M : A
\] The variables occurring in $\Delta$ are considered to be
`boxed': $\Delta$ is a \emph{modal context}. In contrast,
$\Gamma$ is exactly like the run-of-the-mill \emph{intuitionistic
context} we are used to. 

What is \emph{code}? Surely something that does not depend on
normal, \emph{value} variables; only code can contribute to its
making: \[ \begin{prooftree}
  \ctxt{\Delta}{\cdot} \vdash M : A
    \justifies
  \ctxt{\Delta}{\Gamma} \vdash \ibox{M} : \Box M
\end{prooftree} \] This is the introduction rule. As it may be
readily witnessed, it encapsulates an elegant way to enforce the
constraint that code should only depend on code.

The pattern in dual-context systems it that the \emph{elimination
rule} is a \emph{cut rule}. That is, the elimination construct
\emph{is} a very simple explicit substitution. In this particular
case,  it is used to substitute code for a modal variable: \[
\begin{prooftree}
  \ctxt{\Delta}{\Gamma} \vdash \ibox{M} : \Box M
    \qquad
  \ctxt{\Delta, u{:}A}{\Gamma} \vdash N : C
    \justifies
  \ctxt{\Delta}{\Gamma} \vdash \letbox{u}{M}{N} : C
\end{prooftree} \] The reduction rule is the obvious one: \[
  \letbox{u}{\ibox{M}}{N} \longrightarrow
    N[M/u]
\] The rationale is this: the rest of the calculus is responsible
for controlling how modal assumptions are used, and the
elimination rule is used to substitute for one of them. This
constitutes splitting up of the introduction rule of
\cite{Bierman2000a} into two halves, which we then use in the
pattern of introduction and elimination.

What happens to the $\textsf{unbox}$ construct in \emph{op. cit.}
then? The answer is that we hide it in a second \emph{variable
rule}. A modal variable is code; surely we can \emph{evaluate}
this code to obtain a value: \[
  \ctxt{\Delta, u : A, \Delta'}{\Gamma} \vdash u : A
\] This seemingly innocuous `variable rule' is actually a defining
feature of the system: it is the \emph{only} way in which
we can use a modal variable, and it involves \emph{unboxing it}.
Using this elimination rule, rule, one is able to write down a
term of type $\Box A \rightarrow A$, which corresponds to the
\emph{axiom T} of \textsf{CS4}.

In a remarkably lucid sequel, \cite{Davies2001} attempt to justify
their dual-context system using the philosophical approach of
\cite{Martin-Lof1996}, and also include the possibility modality
$\lozenge$, as previously presented by \cite{Bierman2000a} and
\cite{Kobayashi1997}. They also show how one may embed the
\emph{propositional lax logic} \textsf{CL/PLL} in the system, thus
recapturing the \emph{monadic metalanguage} of \cite{Moggi1991}
through their system---see \S \ref{sec:laxlambda}.

To deal with $\lozenge$, a special type of sequent is introduced.
Let us call it a `possibility judgments,' and write
$\ctxt{\Delta}{\Gamma} \vdash M \div A$ for it. Possibility
judgments $\vdash M \div A$ shall be read as `$M$ is a proof that
$A$ is possible.' If $A$ is true, then it is certainly possible;
i.e. normal sequents are included in possibility judgments: \[
  \begin{prooftree}
    \ctxt{\Delta}{\Gamma} \vdash M : A
      \justifies
    \ctxt{\Delta}{\Gamma} \vdash M \div A
  \end{prooftree}
\] This step happens `silently' in \emph{op. cit.} but we prefer
to make it explicit for the sake of clarity. Next, we
\emph{internalise} the possibility judgment, using the $\lozenge$
to stand for it. The introduction rule procees almost
tautologically: if $M$ is possible, then it is true that
$M$ is possible: \[
  \begin{prooftree}
    \ctxt{\Delta}{\Gamma} \vdash M \div A
      \justifies
    \ctxt{\Delta}{\Gamma} \vdash M : \lozenge A
  \end{prooftree}
\] So far we have done nothing more than ensure that truth
implies possibility. The elimination rule completes this picture.
Suppose that there exists a `possible world' of which we know
nothing else, except that $A$ is true. Suppose, furthermore, that
assuming only this fact, that $A$ is true, we can show $C$ to be
possible. Then, if $A$ is possible in this world, then $C$ is
possible in this world as well. In symbols: \[
  \begin{prooftree}
    \ctxt{\Delta}{\Gamma} \vdash M : \lozenge A
      \quad
    \ctxt{\Delta}{x : A } \vdash E \div C
      \justifies
    \ctxt{\Delta}{\Gamma} \vdash 
      \textsf{let dia x } = M \textsf{ in } E : \div C
  \end{prooftree}
\] The associated $\beta$-rule is \[
  \textsf{let dia x } = \textsf{dia } E  \textsf{ in } F
    \rightarrow
  F\langle\langle E / x\rangle\rangle
\] and we shall now define the substitution operator $\langle\langle \cdot /
\cdot \rangle\rangle$.

The reader may think that, surely, we could avoid introducing a
separate type of sequent for possibility. Yet, there is subtlety
that is easy to miss, and which we will use to define the
substitution operator. Notice that if $\ctxt{\Delta}{\Gamma}
\vdash M \div C$ then $M$ can be of one of two forms:
\begin{itemize}
  \item $M$ can be any term; or
  \item $M$ can be of the form 
    $\textsf{let dia x } = M \textsf{ in } E$
\end{itemize} So, let us define a syntactic category of
\emph{proof expressions}, $E$: \[
  E \quad ::= \quad M \;|\; \textsf{let dia } x = M \textsf{ in } E
\] It follows that, since we only use $\langle\langle E /
x\rangle\rangle$ with proof expressions $E$, we can define it by
induction on $E$ rather than by induction on $F$ in
$F\langle\langle E/x\rangle\rangle$. In the case of a term, it
devolves to regular substitution: \[
  F\langle\langle M/x\rangle\rangle := F[M/x]
\] But in the case of a proof expression $\textsf{let dia } x = M
\textsf{ in } E$, we define \[
  F\langle\langle\textsf{let dia } y = M \textsf{ in } E\rangle\rangle := 
  \textsf{let dia } y = M \textsf{ in } F\langle\langle E/x\rangle\rangle
\] This definition unfolds proof expressions in such a way that
fewer commuting conversions are necessary. A similar trick is used
in the $\lambda$-calculus for the lax modality also introduced in
\emph{op. cit.}---see \S \ref{sec:laxlambda}.

The computational interpretation of the diamond modality was
further elaborated by \cite{Pfenning2001}, who demonstrates
that we may think of terms of type $\lozenge A$ as proofs of $A$
whose structure has been hidden.  That is, they are proofs of $A$,
but their internal structure is invisible: they are
\emph{proof-irrelevant}. \cite{Pfenning2001} develops a type
theory based on the juxtaposing the two extremes of this spectrum:
on one hand, terms of box type are only equal up to
$\alpha$-equivalence; on the other hand, terms of diamond type are
all equal. Note that, in order to avoid elimination rules that
look like cut rules, this type theory involves three different
types of $\lambda$-abstraction: one for `boxed' (intensional), one
for intuitionistic, and one for `diamond' (proof-irrelevant)
variables. Pfenning's type theory has not yet been put to the test
of semantical (and/or categorical) analysis.

\cite{Ghani1998} also discuss \textsf{DCS4}, which they use to
repeat the exercise of embedding two-level languages in it, after
suitably extending it with more complicated forms of explicit
substitutions.

\subsection{Categorical Semantics of \textsf{CS4}}

The first to work out the categorical semantics for \textsf{CS4}
seem to be \cite{Bierman1992a, Bierman1996a, Bierman2000a}: they
showed that, to interpret the box fragment of \textsf{CS4}
soundly, one only needs a \emph{cartesian closed category with a
monoidal comonad}.

To interpret the diamond, one needs a \emph{$\Box$-strong monad},
a notion introduced by \cite{Kobayashi1997}:  that is, a monad
$(\lozenge, \eta, \mu)$, along with a $\Box$-strength \[ t_{A, B}:
  \Box A \times \lozenge B \rightarrow \lozenge (\Box A \times B)
\] which is a natural transformation that satisfies appropriate
coherence conditions. The full definition may be found in
\citep{Kobayashi1997} or \citep{Bierman2000a}

However, because of the deficiencies of their term calculus,
Bierman and de Paiva failed to take into account all the necessary
commuting conversions, and hence did not prove a completeness
result. This achivement is again due to \cite{Kobayashi1997}, and
also encompasses his rules for $\lozenge$. Kobayashi also shows
that, were one were to adopt $\eta$-rules for $\textsf{box}$
constructs, then one needs a \emph{cartesian comonad} to achieve
completeness, namely a strong monoidal comonad, which also happens
to be product-preserving between cartesian categories. Further
discussion of categorical semantics for \textsf{CS4} also appears
in \citep{Alechina2001}.

Concrete constructions of models are rare.
\cite{Goubault-Larrecq1999a, Goubault-Larrecq1999b,
Goubault-Larrecq2003} produce some of the few known constructions
of concrete models based on (1) the standard category
$\mathbf{Cpo}$ of complete partial orders (ubiquitously present in
domain theory and most early attempts at semantics for programming
languages); (2) the category $\mathbf{CGHaus}$ of
compactly-generated Hausdorff spaces; and (3) the category of
\emph{augmented simplicial sets}. Of these, the last model is
proven to be complete, in the sense of \cite{Friedman1975}: if two
terms are equal in \emph{every} simplicial augmented model, then
the proof theory equates them as well; see
\citep{Goubault-Larrecq2003}.

A different categorical model for \textsf{CS4}, again on the
category $\mathbf{Cpo}$ of complete partial orders, is presented
in \citep{Kobayashi1997}. The construction is based on a comonad
of \cite{Brookes1991, Brookes1992}. Note, however, that this
comonad also satisfies $A \rightarrow \Box A$ naturally in $A$, so
it is perhaps not the best model to illustrate the computational
behaviour of \textsf{CS4}.

\section{Constructive \textsf{K}}
  \label{sec:ck}

The logic \textsf{CK} can be understood as the smallest normal
modal logic extending intuitionistic propositional logic
(\textsf{IPL}). That is, we only add the axiom \[
  (\mathsf{K)} \quad \Box (A \rightarrow B) \rightarrow (\Box A
  \rightarrow \Box B)
\] in order to obtain its box fragment. To also obtain the
diamond fragment, we throw in \[
  (\lozenge\mathsf{K}) \quad \Box(A \rightarrow B) \rightarrow
    (\lozenge A \rightarrow \lozenge B)
\] A logic very close to \textsf{CK} was the very first
constructive modal logic to be axiomatized, and can almost be
credited to \cite{Wijesekera1990}, but not entirely: Wijesekera's
logic also included $\lozenge \bot \rightarrow \bot$,
which---though adequate for his application---complicates the
analysis of the logic substantially. `Pure \textsf{CK}' itself was
introduced alongside its Curry-Howard interpretation by
\cite{Bellin2001}, and its Kripke semantics were worked out by
\cite{Mendler2005}.

The natural deduction system for \textsf{CK} introduced by
\cite{Bellin2001} is a fragment of a natural deduction system for
classical provability logic (\textsf{GL}), discovered and studied
by one of the authors in the 1980s \citep{Bellin1985}. The central
idea underlying it is the same as those of the Bierman and de
Paiva system for \textsf{CS4}: if the substitution/cut rule is not
admissible, build it in the introduction rule using explicit
substitutions. The difference is that we drop the requirement that
assumptions are boxed. The rule becomes \[ \begin{prooftree}
  \Gamma \vdash M_1 : \Box A_1 
    \quad \dots \quad
  \Gamma \vdash M_n : \Box A_n
    \quad\quad
  x_1 : A_1, \dots, x_n : A_n \vdash N : B
    \justifies
  \Gamma \vdash \textsf{box } N \textsf{ with } M_1, \dots, M_n
  \textsf{ for } x_1, \dots x_n : \Box B
\end{prooftree} \] As one may see, $x_1, \dots x_n$ are available
as `values' within $N$, but we may only place a box in front of
$B$ when all the assumptions going into its making are
`boxed.' Curiously, even though Bellin, de Paiva and Ritter
were the first to seriously propose a proof-theoretic study of
this formulation, it seems that---according to
\cite{Satre1972}---a rule `much like it' was suggested by E.  J.
Lemmon during a lecture in 1966. This rule appears as $\Box I_1$
in \emph{op. cit.} where it is shown that adding it to classical
natural deduction comprises a system deductively equivalent to
classical \textsf{K}.

If explicit substitutions for \textsf{CS4} were problematic, then
the situation for \textsf{CK} is hopeless: the duality between
introduction and elimination ceases to exist altogether, for there
is no elimination rule at all! To match this perfectly, the only
plausible `reduction rule' is very similar to one of the
`secondary' commuting conversions suggested by
\cite{Goubault-Larrecq1996} in the context of \textsf{CS4}. Its
function is to unbox any `canonical' terms in the explicit
substitutions; e.g \begin{align*}
  &\textsf{box } yx \textsf{ with } y, (\textsf{box } M \textsf{
  with } z \textsf{ for } z) \textsf{ for } y, x \\
    \longrightarrow\
  &\textsf{box } y(\textsf{box } M) \textsf{ with } y, z
  \textsf{ for } y, z
\end{align*} in an appropriate context for $y$ and $z$.

In the original presentation of \cite{Bellin2001} there were
multiple technical deficiencies which were remedied by
\cite{Kakutani2007a}. Kakutani's paper also contains many other
gems: a call-by-value version; a CPS transformation from the
call-by-value to the call-by-name version; and soundness and
completeness of categorical semantics. 

In the final part of his paper, \cite{Kakutani2007a} extends the
calculus to \textsf{CK4} and \textsf{CS4} by including
type-indexed constants, \begin{align*}
  &\epsilon_A : \Box A \rightarrow A \\
  &\delta_A : \Box A \rightarrow \Box \Box A
\end{align*} along with some rather complicated equations that
govern their behaviour. Finally, he proves that, were we to
extend the system with the extra structural rules of
\cite{Goubault-Larrecq1996} on explicit substitutions, as in the
case for \textsf{CS4}, one can translate the \textsf{DCS4}
calculus of \cite{Davies2001} to it.

The state of the art for the box fragment of \textsf{CK} is thus
represented by a thoroughly-studied system of explicit
substitutions.

\subsection{Other systems}

There is also another system \textsf{CK}, a dual-context
formulation discussed by \cite{Bellin2001} and \cite{dePaiva2011}.
We do not dwell much on it, as it is seriously pathological (e.g.
weakening is not admissible). \cite{dePaiva2011} also broadly
survey `non-standard' systems for \textsf{CK}, including those in
the style of Fitch, and reiterate the need for a better system.

\subsection{Categorical Semantics of \textsf{CK}}

The categorical semantics of \textsf{CK} were understood by
\cite{Bellin2001} to be a \emph{monoidal endofunctor}, which is
the sole ingredient required to interpret the $(\textsf{K})$
axiom. The proof of soundness and completeness for the system of
\cite{Bellin2001}, in its amended formulation, was provided by
\cite{Kakutani2007a}.

If one wishes to interpret the diamond as well, \cite{Bellin2001}
and \cite{dePaiva2011} remark that we also need a natural
transformation, \[
  st_{A, B} : \Box A \times \lozenge B \rightarrow \lozenge (A
  \times B)
\] This is not as strong as the $\Box$-strong transformation
of \cite{Kobayashi1997} for \textsf{CS4}, nor as strong as the
strong monads of \cite{Moggi1991}. It lies somewhere in between,
and there is no discussion of any coherence conditions it may need
to obey. There is no known proof of soundness or completeness for
this fragment: following the amendments to the calculus presented
by \cite{Kakutani2007a} in order to verify the proof, we consider
the statement in \citep{Bellin2001} wholly unreliable.

\section{Constructive \textsf{K4} \& \textsf{GL}}
  \label{sec:ck4gl}

The only attempt at a natural deduction system for logics such as
\textsf{K4} and \textsf{GL} (Provability Logic) can be found in an
obscure paper of \cite{Bellin1985}. Therein, a system for
\emph{classical} \textsf{GL} is presented, with systems for
\textsf{K} and \textsf{K4} appearing as fragments. We remind the
reader that the axiom \textsf{4}, namely $\Box A \rightarrow \Box
\Box A$, is derivable in provability logic; see, for example,
\cite{Boolos1994} or \cite{Artemov2002}.

Historically speaking, Bellin's system is the first one with
explicit substitutions. One can thus expect that it is not
particularly well-behaved. The situation is worsened by the
inclusion of rules for classical logic, which as is well known
cause problems with normalization. Nevertheless, the paper is very
much concerned with proof theory, and soon the focus of his paper
shifts to the `intuitionistic' fragment of his system.
Unfortunately, the solution he provides is rather complicated, and
we cannot reproduce it in its entirety here.

We can, however, present his rules. The key idea that yields
\textsf{K4} amounts to standing halfway between \textsf{K} and
\textsf{S4}. In term assignment, the appropriate rule would
look like so: \[
  \begin{prooftree}
    \begin{matrix}
      \Gamma \vdash P_1 : \Box A_1 \quad\dots\quad
        \Gamma \vdash P_n : \Box A_n \\
      \Gamma \vdash Q_1 : \Box B_1 \quad\dots\quad
        \Gamma \vdash Q_1 : \Box B_1 \\
      x_1 : \Box A_1, \dots, x_n : \Box A_n,
        y_1 : B_1, \dots, y_m : B_m
          \vdash N : C
    \end{matrix}
      \justifies
    \Gamma \vdash \textsf{box } N \textsf{ with } P_1, \dots, P_n,
    Q_1, \dots, Q_m \textsf{ for } x_1, \dots x_n, y_1, \dots, y_m
    : \Box C
  \end{prooftree}
\] In the Bierman and de Paiva system for \textsf{CS4}, the types
of all the free variables in $N$ had to be `boxed'---i.e. they had
to be of modal type. In the Bellin, de Paiva and Ritter
system for \textsf{CK}, no types of variables at all had to be
boxed---they were available within the (major) premise without a
modality. The case of \textsf{K4} sends us straight to the halfway
house, for we need \emph{both}. But, once more, we have to fulfill
all these assumptions, modal or not, with terms of modal/`boxed'
type.

Why might this be? Intuitively, \textsf{CS4} also had a rule for
`unboxing' things, corresponding to the axiom \textsf{T}.
Therefore, we did not lose anything by requiring that all
assumptions are `boxed,' for we could `unbox` them whenever we
needed that: \[
  x_1 : \Box A_1, \dots, x_n : \Box A_n \vdash (\dots \textsf{unbox }
  x_i \dots) : B
\] In \textsf{K4} this is not available, so we have to make
provisions to accept both `code' and `values.'

How can one adapt this to work for \textsf{GL}, then? The answer
is to make the above rule even more complicated, by adding a
\emph{`boxed' diagonal reference}: \[
  \begin{prooftree}
    \begin{matrix}
      \Gamma \vdash P_1 : \Box A_1 \quad\dots\quad
        \Gamma \vdash P_n : \Box A_n \\
      \Gamma \vdash Q_1 : \Box B_1 \quad\dots\quad
        \Gamma \vdash Q_1 : \Box B_1 \\
      x_1 : \Box A_1, \dots, x_n : \Box A_n,
        y_1 : B_1, \dots, y_m : B_m, z : \Box C
          \vdash N : C
    \end{matrix}
      \justifies
    \Gamma \vdash \textsf{fix } z \textsf{ in }  N \textsf{ with }
      P_1, \dots, P_n, \dots, Q_1, \dots, Q_m
    \textsf{ for } x_1, \dots x_n, y_1, \dots, y_m : \Box C
  \end{prooftree}
\] As in the case of its \textsf{CK}-fragment, this system has no
elimination rule. One $\beta$-rule associated with the above
takes $\textsf{box}$ constructs in the explicit substitutions, and
substitutes them in the term. Another $\beta$-rule defines a
strange recursive procedure, where the entire term is somehow
substituted for occurences of $z$ in itself. This has to be done
very carefully in order to achieve strong normalization; in
particular, some ``doubly diagonal'' assumptions need to be
avoided.

\section{Propositional Lax Logic \textsf{(PLL)}, or Computational
Logic \textsf{(CL)}}

There is a peculiar constructive modality which has arisen in a
multitude of contexts, and thereby repeatedly reinvented.
\cite{Curry1952} discovered its rules and briefly considered it in
the process of trying to prove a deduction theorem for modal
logics. Almost twenty years later, \cite{Goldblatt1981} discovered
a similar modality on topos logic, with the meaning that ``it is
locally the case that.''

The true link with constructivity was made with the investigations of
\cite{Moggi1991} on the \emph{monadic metalanguage}; this included
a modality-like type constructor, which turns out to be logically
equivalent to Curry's modality, as eventually shown by
\cite{Benton1998}, who name it \emph{Computational Logic}
(\textsf{CL}). And then, only a few years later,
\cite{Mendler1993} developed a logic for the verification of
hardware constraits, whose modal propositional fragment turned out
to be coincide with Curry and Moggi. The resulting
\emph{Propositional Lax Logic} (\textsf{PLL}) was studied in
detail by \cite{Fairtlough1995, Fairtlough1997}.

\cite{Fairtlough1995} characterise the logic by
the following axioms: \begin{align*}
  &(\bigcirc R) &A \rightarrow \bigcirc A \\
  &(\bigcirc M) &\bigcirc \bigcirc A \rightarrow \bigcirc A \\
  &(\bigcirc F) &(A \rightarrow B) \rightarrow (\bigcirc A
  \rightarrow \bigcirc B)
\end{align*} Equivalently, \cite{Fairtlough1997} replace axiom
$(\bigcirc F)$ with $\bigcirc A \land \bigcirc \rightarrow
\bigcirc (A \land B)$, which they refer to as $(\bigcirc S)$. One
may wisely ask what modal rules we need. It seems that we do not
need any! Of course, necessitation is improper in this setting.
\cite{Fairtlough1997} suggest a rule which directly follows from
$(\bigcirc F)$ and propositional logic. \cite{Benton1998} suggest
that we need no inference rules related to the modality at all.

Almost immediately, one notices that $(\bigcirc F)$---or
equivalently, $(\bigcirc S)$---is characteristic of necessity
modalities, usually denoted by a box ($\Box$), whereas $(\bigcirc
R)$ and $(\bigcirc M)$ of a possibility modality, commonly denoted
as aa diamond ($\lozenge$). What kind of modality is this, then?
\cite{Fairtlough1997} note that, in a classical context,
$\lnot\bigcirc A$ is true, whether $\bigcirc = \Box$ or $\bigcirc
= \lozenge$. But then, we have that \[
  \bigcirc A \text{ implies }
  \bigcirc \left(A \land \bot\right) \text{ implies }
  \bigcirc A \land \bigcirc \bot \text{ implies }
  \bigcirc \bot \text{ implies }
  \bot \text{ implies }
  A
\] and hence \[
  A \leftrightarrow \bigcirc A
\] That is, $\bigcirc$ cannot be either a classical diamond, or a
classical box.

\subsection{Moggi's calculus}

The natural deduction system for \textsf{PLL/CL} was discovered by
\cite{Moggi1989, Moggi1991}, and it is---at least compared to the other
logics we survey---largely uncontroversial. The introduction
rule simply takes \emph{values} of type $A$ to \emph{computations}
of type $\bigcirc A$: \[ \begin{prooftree}
  \Gamma \vdash M : A
    \justifies
  \Gamma \vdash \textsf{val } M : \bigcirc A
\end{prooftree} \] Under no circumstances should we be allowed to
coerce such a computation back to its previous state as a value.
This is because there are more things of type $\lozenge A$ than
just values of $A$. Moggi's motivation is to use the $\lozenge$
types to isolate a part of the language which \emph{may have side
effects}. Thus, we may also consider \emph{impure} elements of
type $\lozenge A$.

But recall that, in some abstract sense, we know how to
\emph{sequentially compose} side effects. The elimination rule
reflects this intuition: if we can use a value of type $A$ to
produce a computation of type $\bigcirc B$, we may as well plug in
a computation of type $\bigcirc A$ for that value. The resulting
term first performs the side-effects that go into $\bigcirc A$,
and then the ones embedded in the term of type $\bigcirc B$: \[
  \begin{prooftree}
    \Gamma \vdash M : \bigcirc A
      \quad
    \Gamma, x : A \vdash N : \bigcirc B
      \justifies
    \Gamma \vdash \textsf{let val } x = M \textsf{ in } N : \bigcirc B
\end{prooftree} \] The $\beta$-rule is predictable: \[
  \textsf{let val } x = \textsf{val } M \textsf{ in } N
  \longrightarrow N[M/x]
\] The above natural deduction system, as well as equivalent
sequent calculus and Hilbert-style formulations are studied by
\cite{Benton1998}. A sequent calculus is also discussed by
\cite{Fairtlough1997}, but their main focus is on the Kripke
models for \textsf{PLL/CL}. In \citep{Fairtlough2002}, the
inhabitation problem for \textsf{PLL/CL} is discussed, as well as
its interpretation as a logic of `constraint contexts.'
\cite{Goubault-Larrecq2008} discuss a notion of logical relation
appropriate for Moggi's monadic metalanguage.

\subsection{Pfenning \& Davies' lax $\lambda$-calculus}
  \label{sec:laxlambda}

After presenting their dual-context calculus \textsf{DCS4}, for
which see \S \ref{sec:dcs4}, \cite{Davies2001} spend the second
half of their paper discussing
\textsf{PLL/CL}. Using, once more, the quasi-philosophical
approach of \cite{Martin-Lof1996}, they `judgmentally reconstruct'
lax logic, yielding a $\lambda$-calculus that is almost identical
to Moggi's, yet with a subtle distinction in the types of
sequents, some are regular sequents, notated as usual, whereas
some are \emph{lax}, written $\Gamma \vdash E \thicksim A$.
Regular sequents imply lax sequents: \[ \begin{prooftree}
  \Gamma \vdash M : A
    \justifies
  \Gamma \vdash M \thicksim A
\end{prooftree} \] We are only allowed to introduce the modality
only if we have such a lax sequent: \[ \begin{prooftree}
  \Gamma \vdash E \thicksim A
    \justifies
  \Gamma \vdash E : \bigcirc A
\end{prooftree} \] Finally, the elimination rule substitutes a
term of type $\bigcirc A$ into a lax sequent: \[
  \begin{prooftree}
    \Gamma \vdash M : \bigcirc A
      \quad
    \Gamma, x : A \vdash N \thicksim B
      \justifies
    \Gamma \vdash \textsf{let val } x = M \textsf{ in } N \thicksim B
  \end{prooftree}
\] This may seem like a silly game to play, but there is a
subtlety that is easy to miss: well-typed terms $M$ which appear
in a lax sequent $\Gamma \vdash M \thicksim A$ are also elements
of the syntactic category $E$ of \emph{proof expressions}, defined
by \[
  E \quad ::= \quad M\ \;|\;\ \textsf{let val } x = M \textsf{ in } E
\] The importance of this becomes obvious once we glance at the
$\beta$-rule: \[
  \textsf{let val } x = \textsf{val } E \textsf{ in } N
    \longrightarrow
  N \langle E/x \rangle
\] As $E$ is a proof expression, we can define the
substitution operator $N \langle E / x \rangle$ by induction on
$E$, instead of the more common pattern of induction on $N$.
There are two clauses, one for regular terms, and one for
$\textsf{let val}$ constructs: \begin{align*}
  N \langle M/x \rangle &\myeq N[M/x] \\
  N \langle \textsf{let val } y = M \textsf{ in } E
  /x \rangle &\myeq \textsf{let val } y = M \textsf{ in
  } N\langle E/x \rangle
\end{align*} This adjustment to Moggi's system actually removes
the need for its one \emph{commuting conversion}.

\subsection{Categorical semantics of \textsf{PLL/CL}}

The semantics of \textsf{PLL/CLL} is simple and natural: it
consist of a cartesian closed category, with the modality
$\bigcirc$ being the functor part of a \emph{monad}, say
$(\bigcirc, \eta, \mu)$. However, \cite{Moggi1989} remarks this is
not quite enough by itself: it is required that this be a
\emph{strong monad}, in the sense of \cite{Kock1972}. A
\emph{strength}, in this context, is a natural transformation \[
  st_{A, B} : A \times \bigcirc B \rightarrow \bigcirc (A \times B)
\] satisfying some appropriate coherence conditions.

Computationally speaking, the strength corresponds to the ability
to `impurify' any $A$, in order to include it as part of an impure
computation $\bigcirc B$, finally yielding $\bigcirc (A \times
B)$.  Following this setup, \cite{Moggi1989, Moggi1991} obtains
soundness and completeness.

\subsection{The subtle relationship to \textsf{CS4}}

Suppose we have full \textsf{CS4}, diamond modality included.
Then, let us throw in axioms $A \rightarrow \Box A$. Combined with
the \textsf{T} axiom, this trivialises the $\Box$ modality, in the
sense that $A \leftrightarrow \Box A$. However, if we set \[
  \bigcirc \myeq \lozenge
\] then it is not very hard to see that the resulting logic
satisfies the axioms of \textsf{PLL/CL}. 

This is not a mere coincidence. \cite{Alechina2001} showed that,
if one takes Kripke models for \textsf{CS4}, and requires that
they satisfy the semantic counterpart of $A \rightarrow \Box A$
(namely, the accessibility relation is hereditary), then one
obtains sound and complete Kripke semantics for \textsf{PLL/CL}.
It also seems like this semantics coincides with the original
Kripke semantics given in \cite{Fairtlough1997}.

More importantly, this does not appear to be an artifact of the
Kripke semantics: the rabbit hole goes deeper, for the
correspondence is mirrored on the categorical level. To quote
Alechina et al. (with the emphasis being our doing): \begin{quote}
  In the logic, \textsf{PLL} arises as a special case of
  \textsf{CS4} when we assume the derivability of $A \rightarrow
  \Box A$. A similar statement holds in category theory. We have
  an inclusion functor from the category of
  \textsf{PLL}-categories into the category of
  \textsf{CS4}-categories: \emph{each \textsf{PLL}-category is a
  \textsf{CS4}-category where the comonad is the identity
  functor.} Conversely, each \textsf{CS4}-category such that $\Box
  A$ is isomorphic to $A$ is a [sic] \textsf{CS4}-category.
\end{quote} It follows, the authors argue, that $\bigcirc$ really
\emph{is} a possibility modality.

Further evidence of this deep relationship is provided by
\cite{Davies2001}, who embed their lax $\lambda$-calculus into
\textsf{DCS4} (see \S \ref{sec:dcs4}) The key rests in the
following translation of types, where we write $\Rightarrow$ for
implication in \textsf{PLL}: \begin{align*}
  A \Rightarrow B &\myeq \Box A \rightarrow B \\
  \bigcirc A &\myeq \lozenge \Box A
\end{align*} The rules of \textsf{PLL} then become \emph{derivable}
rules of \textsf{CS4}.\footnote{An inference rule is
\emph{derivable} just if, from a proof of the assumptions, a
finite sequence of steps consisting of instances of rules can build a
proof of the conclusion. In contrast, an inference rule is
\emph{admissible} just if, whenever there exists a proof of the
assumptions, then there exists a proof of the conclusion.}

\section{Constructive Linear Temporal Logic (\textsf{CLTL})}
  \label{sec:cltl}

Whilst developing calculi for \textsf{CS4}, \cite{Davies1996,
Davies1999, Davies2001a} noticed that their type systems enforced
an important restriction: their calculi \emph{could not manipulate
code that is not closed}, i.e. code that has free `value'/dynamic
variables, see \cite[\S 8]{Davies2001a}.

That is, if we have a term of shape $\ibox{M}$ at hand, then all
of its free variables are of modal type. This is a very useful
property to know when practicing metaprogramming: we know that
this term only depends on code, i.e. \emph{only depends on static
data}, and we are thus ready to generate code for it. This
generation process may involve some static analysis or local
optimizations applied to $M$, but the calculus shall not control
those: they will happen \emph{once we compile to some other
language}---see e.g. the survey by \cite{Wickline1998}. The
calculus is there only to help us ascertain that things labelled
static depend only on other static data. As a result, \emph{no
unbound dynamic variables will be encountered upon evaluation}.

However, there is a lot of interest in a much more simple and
direct kind of metaprogramming, namely that of \emph{symbolic
execution}. Consider a term $M$ with one a free \emph{dynamic}
variable, $x$. We would like to be able to reduce $M$ whilst
carrying $x$ around as an unevaluated `symbol' until $M$ is
as simple as we would like it to be. A classic example is that of
the exponential function: suppose we have a term $expt$ such that
$expt\ x\ n$ evalutes to $x^n$. We would like some `logical
mechanism' by which we can rewrite $expt$ to some term $expt'$,
and then `partially evaluate' $expt'\ x\ 3$, reducing it to $x * x
* x$. Furthermore, we would like to do this not as a static
analysis on compiled code---as languaged based on \textsf{CS4}
allow us to do---but in a controlled manner \emph{within the modal
calculus}. This is metaprogramming \emph{with open variables}.

To achieve this kind of manipulation, we need to replace the
$\Box$ modality with something which has a more \emph{temporal}
flavour, and which we write $\bigcirc$. This innovation belongs to
\cite{Davies1995, Davies1996a}, who devised a different system,
$\lambda^\bigcirc$. The system itself is a more radical departure
than anything we have discussed up to this point. The main
departure is that both assumptions and sequents are \emph{timed}.

Sequents are of the form \[ \Gamma \vdash^n M : A \] The above
sequent can be read as follows: at time $n$, term $M$
has type $A$, under the assumptions $\Gamma$. But assumptions in
$\Gamma$ are also timed: the assignment $x : A^n$ means that $x$
is a variable of type $A$ at time $n$. One can only use variables
which are available at the `current' time: \[
  \begin{prooftree}
    (x : A^n) \in \Gamma
      \justifies
    \Gamma \vdash^n x : A
  \end{prooftree}
\] Accordingly, we may only $\lambda$-abstract a variable that is
currently available. Moving a time step backwards into the past or
a step forwards into the future is effected by the rules for the
modality $\bigcirc$. If we can infer $A$ in the next moment, then
we can infer that $\bigcirc A$ (`in the next step $A$'): \[
  \begin{prooftree}
    \Gamma \vdash^{n+1} M : A
      \justifies
    \Gamma \vdash^n \textsf{next } M : \bigcirc A
  \end{prooftree}
\] Similarly, if we can infer $\bigcirc A$ at some point, then we can infer
$A$ at the preceding step: \[
  \begin{prooftree}
    \Gamma \vdash^n M : \bigcirc A
      \justifies
    \Gamma \vdash^{n+1} \textsf{prev } M : A
  \end{prooftree}
\] These two constructs give rise to \emph{temporal $\beta$-redices},
namely \[
  \textsf{prev } (\textsf{next } M) \rightarrow M
\] and \emph{temporal $\eta$-redices}, namely \[
  \textsf{next } (\textsf{prev } M) \rightarrow M
\] The natural numbers that decorate sequents carry a very strong
flavour of \emph{time}, and can also be understood as delimiting
\emph{the stage of a computation}. This intuition was formalized
by \cite{Davies1996a}, who proved a \emph{time-ordered
normalization theorem}: suppose $\Gamma \vdash^n M : A$ has no
temporal redices, and no subterm whose derivation is labelled with
time less than $n'$. Then, reducing a $\beta$-redex labelled with
time $n'$ followed by reducing all resultant temporal redices does
not create $\beta$-redices with time less than $n'$. In other
words: reducing a $\beta$-redex of minimal time and then
cancelling $\textsf{prev}$'s with $\textsf{next}$'s will not
produce a redex of lesser time.

\subsection{Logical Aspects}

\cite{Davies1995, Davies1996a} claims to have drawn inspiration for
the design of $\lambda^\bigcirc$ through the Curry-Howard
isomorphism: \begin{quote}
  ``Putting this all together naturally suggests that constructive
  linear-time temporal logic with $\bigcirc$ and a type system for
  multi-level binding-time analysis should be images of each other
  under the Curry-Howard isomorphism.''
\end{quote}

Indeed, in \citep{Davies1996a} it is shown that there exists a term
$M$ such that $\vdash^0 M : A$ if and only if there is a proof of
$A$ in a very small fragment of classical \emph{Linear Temporal Logic
(LTL)}. This fragment is known as $L^\bigcirc$ and is due to
Stirling \cite[p. 516]{Stirling1993}. Axiomatically, $L^\bigcirc$
consists of the normality axiom $\bigcirc(A \rightarrow B)
\rightarrow (\bigcirc A \rightarrow \bigcirc B)$, and the
equivalence of $\bigcirc \lnot A$ and $\lnot \bigcirc A$, along
with modus ponens and necessitation.

However, $L^\bigcirc$ is a classical logic, and hence not an exact
match to the type system of $\lambda^\bigcirc$. The exact logic of
$\lambda^\bigcirc$ is intuitionistic and was isolated by
\cite{Kojima2011}, who call it \emph{Constructive Linear-Time
Temporal Logic (\textsf{CLTL})}. In \emph{op. cit.} the authors study
$\lambda^\bigcirc$ as a natural deduction system, provide an
associated sequent calculus for which they prove cut elimination,
and discuss appropriate Kripke semantics. They also provide an
axiomatization of \textsf{CLTL}, which consists of all instances of
intuitionistic propositional tautologies, as well as the following
axioms: \begin{align*}
  &(\bigcirc K) &\bigcirc(A \rightarrow B) \rightarrow (\bigcirc A
  \rightarrow \bigcirc B) \\
  &(\bigcirc K^{-1}) &(\bigcirc A \rightarrow \bigcirc B)
  \rightarrow \bigcirc (A \rightarrow B)
\end{align*} under the rules of modus ponens and necessitation.

\subsection{Metaprogramming Variants}

Attempts to understand, implement and practice metaprogramming
span multiple decades. Early efforts include \emph{quasiquotation}
in LISP-derived languages, and which date from the mid 1970s
\citep{Bawden1999}; Ershov's idea of \emph{mixed computation}
\citep{Ershov1982}; and the \emph{Futamura projections},
discovered in 1971 \citep{Futamura1999}. The Futamura projections
were later elaborated into the theory and practice \emph{partial
evaluation}; see the survey by \cite{Jones1996} and the book by
\cite{Jones1993}. These developments began to acquire a logical
foundation in the 1990s, mainly through the use of constructive
modalities.

$\lambda^\bigcirc$ was a seminal language in providing this
logical foundation, and it spearheaded many developments. This is
because the constructs $\textsf{next}$ and $\textsf{prev}$
essentially act like the quote and unquote operators of LISP, as
explained in \cite{Bawden1999}.

$\lambda^\bigcirc$ forms the essence of \emph{MetaML}, a
metaprogramming language developed by
\cite{Taha1997, Taha2000}.  Only two fundamental differents
distinguish MetaML to $\lambda^\bigcirc$: \begin{itemize}
  \item MetaML implements \emph{cross-stage persistence}: a
  variable can be used at a \emph{later} stage than its
  annotation, as the variable rule essentially becomes \[
    \begin{prooftree}
      (x :^n A) \in \Gamma
        \quad
      m \geq n
        \justifies
      \Gamma \vdash^m x : A
    \end{prooftree}
  \]

  \item A $\textsf{run}$ construct is forcibly added to the
  system, with reduction rule \[
    \textsf{run}\ (\textsf{next}\ e) \rightarrow e
  \] However, this construct is `unsafe'---in that one can encounter an
  bound runtime variable upon evaluation---and it also comes at a
  cost in `expressivity'; see \cite[\S 12]{Taha2000}.
\end{itemize} MetaML was further simplified in a series of papers
by Moggi, Taha, Sheard and Benaissa.  First, \cite{Moggi1999}
introduced a simplified version called \emph{AIM}, short for
\emph{An Idealized MetaML}. Their work was driven by their attempt
to find a categorical semantics for MetaML, a sketch of which may
be found in an unpublished manuscript \citep{Benaissa1998a}. An
even more simplified language, $\lambda^{BN}$, was introduced by
the same authors a year later \citep{Benaissa1999}, and their
discussion also included a more refined categorical semantics.

A few years later, \cite{Taha2003a} introduced \emph{environment
classifiers}, a system that, amongst other improvements,
constitutes an expansion of $\lambda^\bigcirc$ from \emph{linear}
time to \emph{branching} time. A logical foundation was then
developed for this system by \cite{Tsukada2010}.

Following from the work of Moggi et al, \cite{Yuse2006} introduced
a combination of $\lambda^\bigcirc$ and Davies and Pfenning's
\textsf{DCS4}, called $\lambda^{\bigcirc\Box}$, which combines the
$\bigcirc$ modality of \textsf{CLTL} and the \textsf{CS4} modality
$\Box$. The interpretation is straightforward, and $\bigcirc$ is
supposed to be read as `in the next step' (or stage), whereas
$\Box$ is to be read as `always' (or, in all stages). The
resulting calculus is somehow similar to Moggi et al's \emph{AIM}
and $\lambda^{BN}$, but arguably closer to a more Curry-Howard
based interpretation. Nevertheless, not enough effort is put into
determining which fragment of LTL their calculus corresponds to,
but they do note that it lacks the `induction axiom' $\Box (A
\rightarrow \bigcirc A) \rightarrow (A \rightarrow \Box A$).

As the reader may witness, metaprogramming with modalities has
grown into a vast field of its own that is still expanding. As
such, it will be the main subject of the second part of the
present survey, which Mario Alvarez-Picallo is currently
preparing.

\subsection{Categorical Semantics of \textsf{CLTL}}

\cite{Benaissa1998a} made a start in analyzing the categorical
semantics of various modalities for metaprogramming, and this
included an analysis of $\lambda^\bigcirc$. A seemingly
`conclusive' analysis is given by \cite{Benaissa1999}: a semantics
of \textsf{CLTL} consists of a \emph{cartesian-closed category}
$\mathcal{C}$ and a \emph{full and faithful CCC-functor}, \[
  F : \mathcal{C} \hookrightarrow \mathcal{C}
\] A \emph{CCC-functor} is a finite product preserving (i.e.
strong monoidal) functor for which the canonical arrow $F(B^A)
\rightarrow FB^{FA}$ is an isomorphism.


\end{document}